# Nanosized Sodium-Doped Lanthanum Manganites: Role of the Synthetic Route on their Physical Properties

*Lorenzo Malavasi\*[a], Maria Cristina Mozzati[b], Stefano Polizzi[c], Carlo Bruno Azzoni[b] and Giorgio Flor[a]*

[a]Dipartimento di Chimica Fisica "M. Rolla", INSTM, IENI/CNR Unità di Pavia of Università di Pavia, V.le Taramelli 16, I-27100, Pavia, Italy.
\*E-mail: malavasi@chifis.unipv.it
[b]INFM, Unità di Pavia and Dipartimento di Fisica "A. Volta", Università di Pavia, Via Bassi 6, I-27100, Pavia, Italy.
[c]Dipartimento di Chimica Fisica, Università Ca' Foscari di Venezia, via Torino 155/b, I-30172, Venezia-Mestre, Italy

**Abstract**

In this paper we present the results of the synthesis and characterisation of nanocrystalline $La_{1-x}Na_xMnO_{3+\delta}$ samples. Two synthetic routes were employed: polyacrylamide-based sol-gel and propellant synthesis. Pure, single phase materials were obtained with grain size around 35 nm for the sol-gel samples and around 55 nm for the propellant ones, which moreover present a more broaden grain size distribution. For both series a superparamagnetic behaviour was evidenced by means of magnetisation and EPR measurements with peculiar features ascribable to the different grain sizes and morphology. Preliminary magnetoresistivity measurements show enhanced low-field (< 1 T) magnetoresistance values which suggest an interesting applicative use of these manganites.

# Introduction

Presently, one of the most exciting research field in rare earth manganite perovskites concerns with the study of nano-structured materials in order to induce new and peculiar phenomena such as the low-field magnetoresistance (LFMR). LFMR consists in the resistivity variation in response to very low applied magnetic fields with respect to those normally needed for the insurgence of colossal magnetoresistance (CMR); LFMR usually ranges from few percent to few tens percent for applied magnetic fields lower than 1 T [1-3].

Even though several reports about nanocrystalline manganites are present in the current literature, the debate on the effective transport mechanism involved in the magnetoresistive properties of these materials is still open since the usual Zener's double exchange picture, valid for the bulk manganites, does not hold in the nanometric materials for which a mechanism based on the spin-polarised tunneling of charge carrier between grains has been invoked [4-6]. One of the fundamental pre-requisite for rationalising and comparing the results on nanocrystalline manganites and, in general, on nanosized materials, is a careful and thorough characterisation of the synthesis products. This is a key factor since the effective morphology, the chemical composition and the grain size distribution deeply affect the physico-chemical properties of the materials prepared by the various currently preparative routes for obtaining materials with grains of the order of few tens of nanometers [7-10].

In this paper we focused on the synthesis of a series of $La_{1-x}Na_xMnO_{3+\delta}$ samples with $x = 0$, 0.15 and 0.2 carried out by means of a sol-gel technique, derived from the Pechini method [11], and propellant synthesis. The choice of this dopant with respect to Ca or Sr comes from the observation that the ionic radius of sodium is closer to the one of lanthanum so the tolerance factor ($t$) is practically unchanged by the substitution; moreover, compared to the Ca or Sr-doped manganites, it is possible to achieve an equal amount of hole doping with a lower cation substitution since for the same amount of aliovalent dopant the hole density is twice with respect to the calcium or strontium doping. This should reflect in a lower cation disorder induced by the doping. Finally, it was observed that with the Na-doping it is possible to obtain high values of magnetoresistivity (MR), as for the Ca-doped samples, but for higher temperatures, closer to room temperature. For the Sr-doped manganites the transition temperatures are higher but with low MR values.

The cation stoichiometry of the prepared materials has been carefully checked by means of Micro-probe analysis. X-ray powder diffraction investigation (XRPD) was used to reveal the structural features of the two series of samples, Transmission Electron Microscopy (TEM) have been combined in order to investigate their morphology. Static magnetization and Electron Paramagnetic Resonance (EPR) measurements were used to study the magnetic properties which have been correlated to the structural and morphological features of the nanopowders resulting from the two synthetic routes. Preliminary magnetoresistivity (MR) measurements were also performed. To our knowledge this is the first study devoted to the synthesis and characterisation of sodium-doped nanocrystalline manganites.

## Experimental

$La_{1-x}Na_xMnO_{3+\delta}$ samples with nominal $x$ values of 0, 0.15 and 0.2 have been prepared by means of sol-gel and propellant synthesis.

For the sol-gel synthesis, manganese nitrate (Aldrich 99.99+%), lanthanum nitrate (Aldrich 99.99+%) and sodium nitrate (Aldrich 99.99+%) were dissolved separately in a small quantity of deionised water and a stoichiometric amount of acid-EDTA ($C_{10}H_{16}N_2O_8$, Ethylenediamintetraacetic acid, Aldrich 99.5%) was added in the ratio 1:1 with continuous stirring. The solution was heated between 333 K and 353 K and the pH was adjusted with ammonium hydroxide (Aldrich, 99%) in order to allow the formation of a complex. When the solution was clear the monomers acrylamide, $H_2C=CHCONH_2$, (Aldrich 99+%) and *N,N'*-methylenebisacrylamide, $H_2C=CHCONH)_2CH_2$, were added to the solution. The ratio of crosslinker, the *N,N'*-methylenebisacrylamide, and the amount of gel affect the grain size of the prepared powder. Here we used 6 grams of Acrylamide and 1 gram of *N,N'*-Methylenebisacrylamide in 100 ml of solution [7]. A small amount of $\alpha,\alpha'$-azoisobutyronitrile (AIBN), $C_8H_{12}N_4$ (Fluka $\geq$ 98.0%), was added in order to activate the polymerisation process. The solution was then slowly heated with continuous stirring until 353 K when the gel formed. The gel was dried in a microwave oven and homogenised in a mortar. The xerogel was calcinated at 923 K and at 1173 K for 60 minutes.

For the propellant synthesis proper amounts of manganese nitrate (Aldrich 99.99+%), lanthanum nitrate (Aldrich 99.99+%) and sodium nitrate (Aldrich 99.99+%) were used as oxidizers while urea (Carbonyldiamide, $CH_4N_2O$, Fluka >99.5%) was used as fuel. Oxidizers and fuel were dissolved in a small quantity of deionised water and heated on a hot-plate. The amount of urea was calculated in order to obtain an oxidizer/fuel ratio equal to 1 according to the following equation:

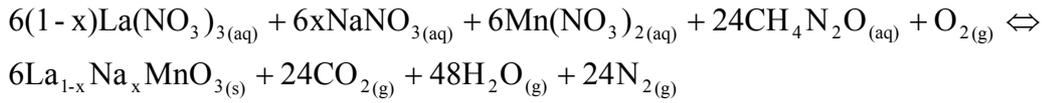

$$6(1-x)La(NO_3)_{3(aq)} + 6xNaNO_{3(aq)} + 6Mn(NO_3)_{2(aq)} + 24CH_4N_2O_{(aq)} + O_{2(g)} \Leftrightarrow$$
$$6La_{1-x}Na_xMnO_{3(s)} + 24CO_{2(g)} + 48H_2O_{(g)} + 24N_{2(g)}$$

When a critical temperature was reached (around 573 K) the solution boiled, frothed, turned dark and then ignited and caught fire to give a spongy powder. The as-prepared powder was calcined in a platinum crucible at 973 K for one hour.

XRPD patterns were acquired on a Philips 1710 diffractometer equipped with a Cu anticathode, adjustable divergence slit, graphite monochromator on the diffracted beam and a proportional detector. Lattice constants were determined by minimizing the weighted squared difference between calculated and experimental $Q_i$ values, where $Q_i = 4sin^2\theta_i/\lambda_i^2$ and weight = $sin(2\theta_i)^{-2}$. Instrumental aberrations were considered by inserting additional terms into the linear least square-fitting model [12].

The well-known Scherrer formula was applied to determine the average grain size:

$$D = \frac{0.9 \cdot \lambda}{\cos(2\vartheta) \cdot L} \tag{1}$$

where D is the grain size expressed in nanometers, $\lambda$ is the wavelength of the incident X-rays, i.e. 0.154056 nm, $2\vartheta$ is the peak position estimated by a fitting procedure with the Philips APD software which also allowed the determination of the full width at peak half maximum (L) expressed in radians and corrected for the instrumental broadening by measuring the width of a standard reference.

Electron microprobe analysis (EMPA) measurements were carried out using an ARL SEMQ scanning electron microscope, performing at least 10 measurements in different regions of each sample. According to EMPA and XRPD data, the above synthetic procedures gave single-phase materials; in addition each sample was found to be highly homogeneous in the chemical composition, which was met in fair agreement with the nominal one.

TEM images were taken at 300kV using a Jeol 3010 with a high resolution pole piece (0.17 nm point-to-point resolution) and equipped with a Gatan slow-scan 794 CCD camera. The powder was dispersed in isopropyl alcohol and sonicated for 5 min. A 5 μL drop of the solution was deposited onto a carbon holey film.

Magnetisation measurements were carried out with a SQUID magnetometer by applying different magnetic fields (0 - 7 T) in the temperature range 2 - 350 K. MR measurements were carried out between 300 and 4.2 K at various fields with the DC-four electrodes method by means of a specific probe directly inside the SQUID apparatus. To avoid grain growth during sintering step the powders were pressed, by means of an isostatic press, in form of pellets and fired at 923 K (Sol-gel synthesized sample) and 973 K (Propellant synthesized sample) for 1 hour, respectively.

EPR measurements in X band (~ 9.5 GHz) were performed using a Bruker spectrometer, with a continuous nitrogen flow apparatus, in the range 120-450 K.

**Results and Discussion**

*1- X-ray diffraction and Transmission Electron Microscopy*

Figure 1 reports the XRPD patterns of the $La_{1-x}Na_xMnO_{3+\delta}$ samples for nominal *x* values of 0 (b), 0.15 (c) and 0.2 (d), obtained by the sol-gel synthesis and heated at 923 K for 1 hour . The XRPD pattern of pure $LaMnO_{3+\delta}$ heated for 1 hour at 873 K (a) is also shown. Vertical lines in the Figure refer to the reference pattern for the rhombohedral manganite [13]. The asterisk in the Figure refers to the peak of the sample holder which is made of aluminium. As it can be appreciated, the samples heated at 923 K display patterns compatible with well crystallised, single-phase materials with a rhombohedral perovskite structure (Space Group 167). On the comparison it is possible to note that a thermal treatment at 873 K for 1 hour (a) does not allow the formation of a pure, single-phase material since the amorphous band is still present.

Lattice constants for the three sol-gel synthesised (S-G) samples are reported in Table 1. The chemical composition of these samples has been carefully checked by means of EMPA measurements. We could observe a fairly good agreement between the nominal chemical composition and the effective one, which however presents a larger sodium content. The actual stoichiometries will be hereafter used instead of the nominal ones.

Lattice constants show a small increase by increasing the sodium doping even though the differences between the three samples are small, as it can be appreciated by comparing the cell volumes. This increase is probably due to the difference in the ionic radii of $La^{3+}$ and $Na^{+}$. This trend also suggests that the $Mn^{3+}/Mn^{4+}$ ratio of these samples is more or less the same, otherwise a general decrease of cell volume should be observed. Let's note that this ratio should be the most important variable in affecting the cell parameters since the ionic radii difference between the two manganese ions is larger than that between $La^{+3}$ and $Na^{+}$ ions [14]. The oxygen content of the samples was checked by means of chemical reduction at 1073 K with a mixture of hydrogen (5%) in argon. The results show that pure LaMnO$_{3+\delta}$ is oxygen over-stoichiometric (0.18±0.01), while for $x = 0.17$ a nearly correct oxygen content was found; for $x = 0.24$ the sample displays an oxygen under-stoichiometry of about 0.05. These results are in agreement with the general behaviour of this class of perovskites with respect to the variation of the oxygen content along with the cation doping [15-17]; in particular it was established that in air at temperature around 973 K the oxidation state of different series of doped-manganites has a common and mean value around 33-35% and that the effect of dopant content increase is a decrease of the oxygen content [18-20].

At this point we should note that the presence of oxygen over-stoichiometry ($\delta > 0$) is not properly accounted for in the formalism LaMnO$_{3+\delta}$ since it is well established that the extra-oxygen is not of interstitial-type but rather cation vacancies are formed for compensation [21, 22]. So, the correct formula for these compounds should be: La$_{1-\varepsilon}$Mn$_{1-\varepsilon}$O$_3$ where $\varepsilon = \delta/(3+\delta)$. Anyway, this notation, although correct, is less straightforward than the one usually adopted in the current scientific literature regarding manganites, *i.e.* LaMnO$_{3+\delta}$. As a consequence we are going to use this last formalism having in mind, anyway, the correct defect chemistry of oxygen over-stoichiometric materials.

An analogous X-ray characterisation has been carried out for the samples prepared by propellant synthesis and heated at 973 K for 1 hour. The results are presented in Figure 2 for the samples with nominal *x*-values of 0 (c), 0.15 (d) and 0.2 (e). In this Figure the X-ray patterns of the propellant products heated at 873 K (a) and at 923 K (b) and the reference lines for the rhombohedral perovskite [13] are also shown. The corresponding lattice parameters are reported in Table 1 together with the actual chemical composition as deduced from EMPA measurements.

From the diffraction patterns of the propellant synthesised (PR) samples it is possible to conclude that the formation of single phase, well crystallised, rhombohedral manganites is achieved at 973 K while lower temperatures do not lead to the formation of pure crystalline materials.

From Table 1 it is possible to appreciate that also in this case the differences in the lattice constants of the three samples are very small, again suggesting the presence of similar Mn$^{+3}$/Mn$^{+4}$ ratios. Besides the reduction experiments gave results analogous to those obtained for the S-G samples (the oxygen content values are also reported in Table 1). So, within the experimental error, we can say that the $Mn^{+4}$ content of the prepared samples is nearly constant along the solid solution. Moreover, the lattice parameters of the two series of samples (S-G and PR) are quite similar even though, for the same composition, the cell volumes for the PR samples display smaller volumes. This result may be correlated to the different grain dimensions of the samples of the two series (see later in the text) as analogously found in some other oxides systems [23].

The average grain dimension of all the samples was estimated by means of the Scherrer formula (see the Experimental part for details). The results are presented in Table 1. The grain dimension of the S-G nanopowders is smaller than that of the PR samples by about 20 nm. Let's note that this can be partially connected to the fact that the lowest temperature needed to obtain a monophasic compound is 923 K for the sol-gel procedure and 973 K for the propellant one.

A clearer picture of these features has been developed by performing TEM analysis. Figure 3 shows selected TEM images recorded for S-G and PR $LaMnO_{3+\delta}$. This Figure allows to appreciate that the S-G material is composed of separated grains whose size is of the order of 30 nm, in accordance with the X-ray determination. On the other hand, the TEM inspection for the PR samples, reported in Figures 3b, reveals a more agglomerated structure of the nanopowders and a wider grain size distribution with respect to the S-G sample with grain sizes ranging from about 20 nm to about 100 nm.

*2- Static magnetization, Electron Paramagnetic Resonance, magnetoresistance*

Magnetization (*M*) measurements *vs.* temperature have been carried out by cooling the samples both without applying a magnetic field (zero field cooling, ZFC) and by applying the measurement field (field cooling, FC). Figure 4 reports the FC molar *M(T)* curves acquired at 1000 Oe for the S-G (open symbols) and for the PR (full symbols) synthesized samples. In the inset, it is shown the difference between ZFC and FC curves at 1000 Oe for one representative sample of each series (*i.e.* S-G La$_{0.83}$Na$_{0.17}$MnO$_{3+\delta}$ and PR La$_{0.84}$Na$_{0.16}$MnO$_{3+\delta}$), which suggests the presence of magnetic domains.

In both the undoped samples (S-G and PR LaMnO$_{3+\delta}$) the expected antiferromagnetic transition at about 140 K does not occur due to the oxygen overstoichiometry [24]. The great amount of cation vacancies introduced according to the following equilibrium:

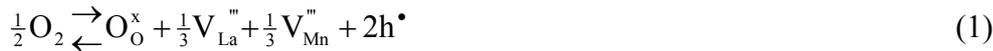  (1)

$$\tfrac{1}{2}O_2 \rightleftarrows O_O^x + \tfrac{1}{3}V_{La}''' + \tfrac{1}{3}V_{Mn}''' + 2h^\bullet$$

is possibly responsible for the very low *M* values observed.

Concerning the S-G Na-doped samples, it is mainly noticeable that the *M*(*T*) curve shape is not the common one encountered in polycrystalline ferromagnetic materials [25]: no sharp para-ferromagnetic (PF) transition is present while a slow enhancement of the magnetization by lowering the temperature is observed. Similar behavior concerned with analogous bulk samples for which the broadness of the magnetic PF transition was ascribed to the presence of separated samples regions, with different $T_c$ values, as also suggested by EPR and electrical results [16]. For these S-G samples the role of the nanosized grains on the magnetic properties (the grain size reduction may lead to partially fluctuating spin systems) is better outlined by the *M vs. H* measurements.

Figure 5a reports the *M vs. H* curves at 300, 180, 130, 30 and 2 K for the S-G La$_{0.83}$Na$_{0.17}$MnO$_{3+\delta}$ sample, selected as an example. Only at 300 K a pure paramagnetic behavior is evidenced. As temperature decreases the curves try to reach more rapidly their asymptotic values but the magnetization never saturates. This feature strongly suggests the presence of superparamagnetism. Besides, the magnetization achieved at 30 K and 2 K is nearly the same as a further proof of an inability of the system to reach a complete magnetic order; moreover the maximum reached value is much lower than the expected one for bulk compounds [25]. On the other hand magnetization saturation is observed on the same S-G sample annealed at 1173 K for 1 hour, for which we found a grain size of about 200 nm. Figure 6 shows the results of the *M vs. H* investigation for this sample: the magnetization saturation is now reached at 30 K and 1.5 T similarly to bulk materials, with saturation value close to the expected one for an analogous $Mn^{+4}$ content [25].

From Fig. 4 it is evident that also for the PR samples no sharp PF transitions are observable: again, the *M* slowly increases by reducing the temperature but with a different trend with respect to the S-G samples. For the PR Na-doped samples the magnetization increases more resolutely and at higher temperatures with respect to the S-G ones; this is more similar to bulk systems and is in agreement with the higher grain dimensions we observed. Besides, for lower *T*, the curves keep on increasing very slowly up to the lowest temperatures; this trend may be correlated to the larger grain size distribution and inhomogeneity evidenced by the morphologic characterization.

*M vs. H* measurements, performed at 350, 300, 180, 30 and 2 K, are presented in Figure 5b for the $La_{0.84}Na_{0.16}MnO_{3+\delta}$ PR sample, selected as an example. This sample shows characteristics very similar to those observed for the S-G one, so that also in this case a superparamagnetic behavior is outlined. The achieved *M* value at 2 and 30 K is only slightly smaller with respect to the previous sample, and again far from the expected bulk value.

The above data analysis evidences, for all the investigated samples, the typical phenomenological behavior of superparamagnetic systems, with a gradual *M* enhancement for the S-G samples and a more drastic *M* enhancement for the PR ones, at least at the highest temperatures. However, due to the great amount of variables pertaining to these compounds, a simple Langevin function does not allow a satisfactory reproduction of the *M*(*T*) and *M*(*H*) curves. For instance, the magnetic field increase strongly affects the single grain magnetization, inducing the system to enhance its order. Moreover, the temperature dependence of single grain magnetization is rather complicated since it is strictly related to the grain volume and homogeneity of the samples. Additional difficulties concern with the interpretation of the PR samples, for which also a grain size and $T_c$ distribution should be considered.

More detailed information on the magnetic characteristics of these samples, related to their magnetic and chemical homogeneity, can be argued from the EPR data, which also allow to evidence the main magnetic differences from the bulk compounds.

In Fig. 7 the EPR signals of two representative samples (S-G $La_{0.76}Na_{0.24}MnO_{3+\delta}$ and PR $La_{0.76}Na_{0.24}MnO_{3+\delta}$) are reported at different temperatures. For the S-G samples (see Fig7a) only one signal is observed with g $\cong$ 2 for $T$ higher than a peculiar temperature, defined as $T_{onset}$, thus suggesting the whole paramagnetic character of the material. For $T < T_{onset}$ only one signal is observed with $g_{eff} > 2$. This value progressively increases by lowering the temperature, thus evidencing the progressive extent of internal magnetic fields. At the same time, the signal broadens by lowering the temperature. The presence of only one signal over the whole investigated temperature range, in particular for $T < T_{onset}$, suggests that a good magnetic and chemical homogeneity characterizes the S-G samples. So the slow enhancement of the static magnetization by lowering the temperature (see Fig. 4) can be strictly related to the superparamagnetic properties of these materials. The EPR spectrum of the S-G samples annealed at 1173 K (not shown) clearly shows two different components for $T < T_{onset}$, indicating that the annealed samples loose the high homogeneity and look more similar to the bulk compounds, for which two signals with different $T_{onset}$ were previously observed [16]. In this case the coexistence of sample regions with different grain dimensions (micro- and nano-metric, possibly with different $T_c$ values), induced by the high temperature annealing, can be responsible for the two observed components of the spectrum.

The main features pertaining to the S-G samples also characterize the spectrum of the PR samples, as shown in Fig. 7b. However in this case a very weak second component, mostly evident in the undoped sample, can be responsible for the different intensity behavior for $T < T_{onset}$ and can justify the different $M(T)$ behavior of the PR samples with respect to the S-G ones.

The signal intensity temperature dependence in the proximity of $T_{onset}$, reflects the $M(T)$ curve behavior for both sample series. A slow, gradual enhancement is observed for the S-G samples, whereas a more drastic intensity enhancement characterizes the PR samples, suggesting their higher rapidity to form clusters of oriented spin. This reminds of the anomalous intensity enhancement previously observed in this $T$ range for bulk compounds [16, 26, 27], so that the lack of saturation in the $M(H)$ curves represents the main difference between PR nanopowders and bulk compounds.

Some details concerning the superparamagnetic characteristics of these systems can be argued by analyzing the relationship between the EPR resonant field shift ($\delta H_r$) and the line-width ($\Delta H$) for $T < T_{onset}$. Figure 8 shows $\delta H_r$ *vs.* $\Delta H$ for the samples represented in Fig. 7. A cubic dependence is observed for the S-G samples whereas a quadratic or cubic one is suitable for the PR compounds. The dependence observed for both samples is typical of superparamagnetic systems, as empirically studied for ultrafine magnetic particles [28]. Moreover, some suggestions about the particles orientation can be obtained from these curves, even if inter-grain magnetic interactions should be also considered in our samples. In particular, the cubic dependence suggests the presence of randomly oriented particles, while partial particles orientation should give rise to a quadratic dependence. Again clearer superparamagnetic characteristics pertain to the S-G samples, consistently with their lower grain size, so with the most important role of the surface and with a higher fluctuation of the spin structure.

On the two series of samples we also carried out preliminary magnetoresisitivity (MR) measurements. In this case we just wanted to check the insurgence of the new properties connected with the grain dimension reduction also for what concerns the transport properties. We performed the MR measurements on S-G $La_{0.83}Na_{0.17}MnO_{3+\delta}$ and on PR $La_{0.84}Na_{0.16}MnO_{3+\delta}$ samples, that is the same of the inset of Figure 4.

Figure 9 reports the $\rho(T)$ and MR curves for the S-G sample at 0, 0.01, 0.1, 1 and 5 Tesla. The resistivity curve at zero field presents a transition from a semiconducting-like to a metallic-like regime (S-M) at about 176 K (taken at the maximum of the curve) which agrees with the inflection point on the $M(T)$ curve (see Figure 4), where a conspicuous fraction of the sample is ferromagnetically oriented, thus favoring the insurgence of the metallic regime. The curve also displays a rise of the resistance at low temperatures, which can be accounted for the presence of grain boundaries in the material [29] as understandable due to the low temperature of the firing. By applying a magnetic field the resistivity decreases. For the lowest applied magnetic field (0.01 T) it is possible to note a MR effect of more than 10% for temperatures lower than 100 K; by increasing the temperature the MR effect progressively disappears. At 0.1 T, which anyway is a *low* magnetic field if compared to the ones required to the insurgence of the CMR effect, a MR effect of about 20 % is already achieved at 100 K. Under both these fields the behavior of the resistance is typical of nanosized materials, that is, characterized by a smooth increase of the MR effect as the temperature decreases [30]. This behavior is different with respect to the CMR where the maximum value of MR is achieved in correspondence of the S-M transition temperature as happens for the highest investigated field (5 T). At 1 T, which is considered the reference value for the so called LFMR, our sample has, at the lowest temperature explored, a MR value of about 40% which is among the highest reported in the literature till now [31-32].

For the PR $La_{0.84}Na_{0.16}MnO_{3+\delta}$ sample the corresponding $\rho(T)$ and MR curves at 0, 0.5 and 5 T are presented in Figure 10. In this case the MR values are lower with respect to the previous sample. At 0.5 T the MR reaches a value around 10% at 4 K with again a smooth decrease as the temperature rises. Interestingly, at 5 T the MR behaves as in 0.5 T, *i.e.* without a MR peak around the S-M transition. This is in agreement with literature reports which indicate that as grain size reduces also the High-Field Magnetoresistance (HFMR) increases [30] (note that the average grain sizes are 40 nm and 62 nm for the S-G and PR samples, respectively).

**Concluding remarks**

In the present work we have carried out the synthesis of nanosized sodium-doped lanthanum manganites (La$_{1-x}$Na$_x$MnO$_{3+\delta}$) by means of two distinct chemical routes: polyacrylamide-based sol-gel and propellant synthesis. The amorphous as-prepared materials underwent different thermal treatments thus showing that the lower required temperature to obtain a crystalline, single phase oxide is 923 K with the sol-gel route and 973 K with the propellant one.

The morphology of the samples prepared by means of the two techniques is rather different: for the S-G prepared materials a smaller grain size distribution and dimension was achieved with respect to the PR samples. Even though the grain size difference ranges around 20 nm it significantly affects the magnetic behaviour.

According to magnetisation and EPR measurements both samples series clearly revealed a superparamagnetic behaviour, particularly evidenced by the lack of magnetisation saturation also for the lowest investigated temperature. Indeed the magnetisation does not reach the final value expected for a collective long range spin order but gradually increases while the system is in a local scale magnetically ordered state. The different origin of superparamagnetism in the two series is revealed by the different magnetic behaviour *versus* temperature, as pointed out by the two techniques. Depending on the different grain sizes and distribution, a purer superparamagnetic behaviour characterises the S-G samples, as also evidenced from the EPR resonant field shift dependence on the line width. The overall collected data show that the S-G synthesis allows a better control of the grain size distribution with a relatively fast and low-cost process. The preliminary magnetoresistivity (MR) data collected show an interesting behaviour for both the samples. The S-G ones, anyway, present a MR value at 1 T which, at present, is one of the highest reported for alkali or alkali-earth doped LaMnO$_3$ manganites. This result makes this material highly promising for applicative purposes. Further investigations will point towards the way to increase the MR values and in particular to extend the temperature range of the MR phenomenon at higher temperature.

## Acknowledgments

Cedric Philereau is gratefully acknowledged for samples preparation and Dr. Simona Bigi for having performed EMPA analysis. The Department of Earth Science of Modena University and CNR of Modena are acknowledged for allowing EMPA use. E. Jarosewich has kindly supplied the lanthanum standard for the EMPA measurements. Dr. Oscar Barlascini is acknowledged for having performed the EPR measurements. Financial support from the Italian Ministry of Scientific Research (MIUR) by PRIN Projects (2002) is gratefully acknowledged.

**Figure captions**

**Fig. 1.** X-ray powder patterns for $La_{1-x}Na_xMnO_{3+\delta}$ samples prepared by sol-gel with nominal $x$ values of 0 (b), 0.15 (c) and 0.2 (d) heated at 923 K for 1 h. Pattern (a) refers to the sample with $x = 0$ heated at 873 K. Vertical lines represents the reference pattern for rhombohedral perovskites. The asterisk refers to the peak of the sample holder made of aluminium.

**Fig. 2.** X-ray powder patterns for $La_{1-x}Na_xMnO_{3+\delta}$ samples prepared by propellant synthesis with nominal $x$ values of 0 (c), 0.15 (d) and 0.2 (e) heated at 973 K for 1 h. Patterns (a) and (b) refer to the sample with $x = 0$ heated at 873 K and 923 K respectively. Vertical lines represent the reference pattern for rhombohedral perovskites. The asterisk refers to the peak of the sample holder made of aluminium.

**Fig. 3.** TEM images for S-G $LaMnO_{3+\delta}$ (3a) and for PR $LaMnO_{3+\delta}$ (3b). Reference lines in the Figure are 20 nm for both samples.

**Fig. 4.** FC molar magnetisation at 1000 Oe for the S-G samples (open symbols) with $x = 0$ (square), $x = 0.17$ (circles) and $x = 0.24$ (triangles) and PR samples (full symbols) with $x = 0$ (square), $x = 0.16$ (circles) and $x = 0.23$ (triangles). In the inset the FC and ZFC curves at 1000 Oe for S-G $La_{0.83}Na_{0.17}MnO_{3+\delta}$ (open circles) and PR $La_{0.84}Na_{0.16}MnO_{3+\delta}$ (full circles) are reported.

**Fig. 5.** *M vs H* curves at different temperatures for S-G $La_{0.83}Na_{0.17}MnO_{3+\delta}$ (a) and PR $La_{0.84}Na_{0.16}MnO_{3+\delta}$ (b) samples presented as molar magnetization (left axis) and magnetic moment per unit formula (u.f.) in Bohr magnetons ($\mu_B$) units (right axis).

**Fig. 6.** *M vs H* curves at different temperatures for the same sample of Figure 5a fired at 1123 K presented as molar magnetization (left axis) and magnetic moment per unit formula (u.f.) in Bohr magnetons ($\mu_B$) units (right axis).

**Fig. 7.** EPR signals at different temperatures for S-G $La_{0.76}Na_{0.24}MnO_{3+\delta}$ (a) and PR $La_{0.77}Na_{0.23}MnO_{3+\delta}$ (b) samples.

**Fig. 8.** $\delta H_r$ *vs* $\Delta H$ for S-G La$_{0.76}$Na$_{0.24}$MnO$_{3+\delta}$ (a) and PR La$_{0.77}$Na$_{0.23}$MnO$_{3+\delta}$ (b) samples. Solid and dashed lines represent the cubic and quadratic dependence, respectively. Symbols represent the experimental values.

**Fig. 9.** Resistivity behaviour (a) and MR values (b) for S-G La$_{0.83}$Na$_{0.17}$MnO$_{3+\delta}$ sample.

**Fig. 10.** Resistivity behaviour and MR values for PR La$_{0.84}$Na$_{0.16}$MnO$_{3+\delta}$ sample.

**Table caption**

**Table 1.** Effective sodium content (*x*), lattice parameters (*a* = *b* and *c*), cell volumes (*V*) and average grain dimension (*D*) estimated from the Scherrer formula for the samples considered in the paper.

**Table 1**

| Sample | *Effective x* | *a* (Å) | *c* (Å) | *V* (Å$^3$) | *D* (nm) |
|---|---|---|---|---|---|
| LaMnO$_{3.18}$ (S-G) | 0 | 5.501 | 13.3134 | 58.147 | 30 |
| La$_{0.85}$Na$_{0.15}$MnO$_3$ (S-G) | 0.17 | 5.5046 | 13.3267 | 58.282 | 40 |
| La$_{0.8}$Na$_{0.2}$MnO$_{2.95}$ (S-G) | 0.24 | 5.510 | 13.3409 | 58.459 | 36 |
| LaMnO$_{3.175}$ (PR) | 0 | 5.499 | 13.3102 | 58.092 | 55 |
| La$_{0.85}$Na$_{0.15}$MnO$_3$ (PR) | 0.16 | 5.4934 | 13.3357 | 58.085 | 62 |
| La$_{0.8}$Na$_{0.2}$MnO$_{2.96}$ (PR) | 0.23 | 5.4981 | 13.3389 | 58.198 | 55 |

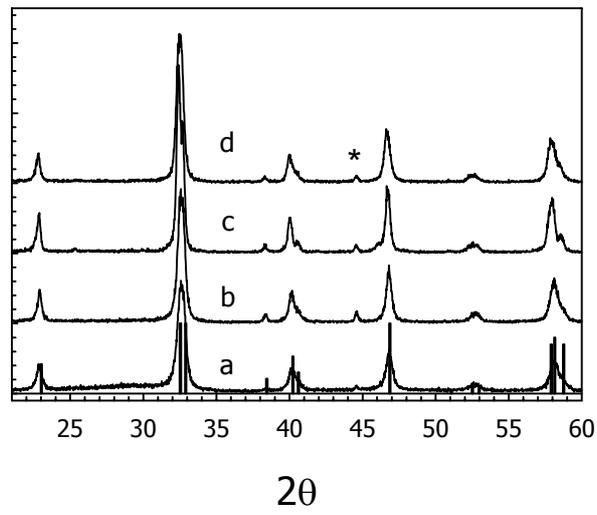

Figure 1

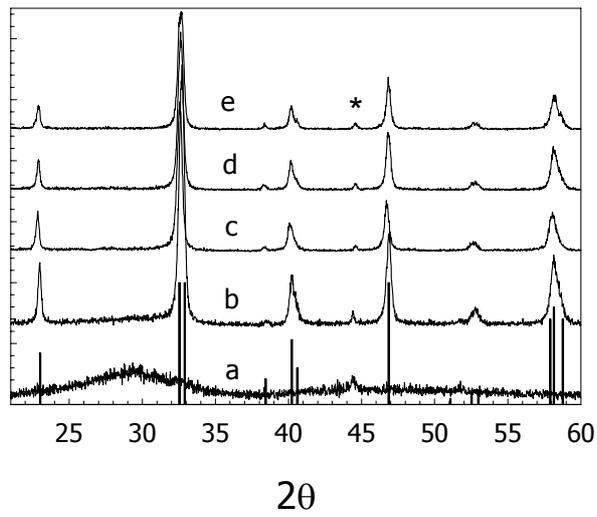

Figure 2

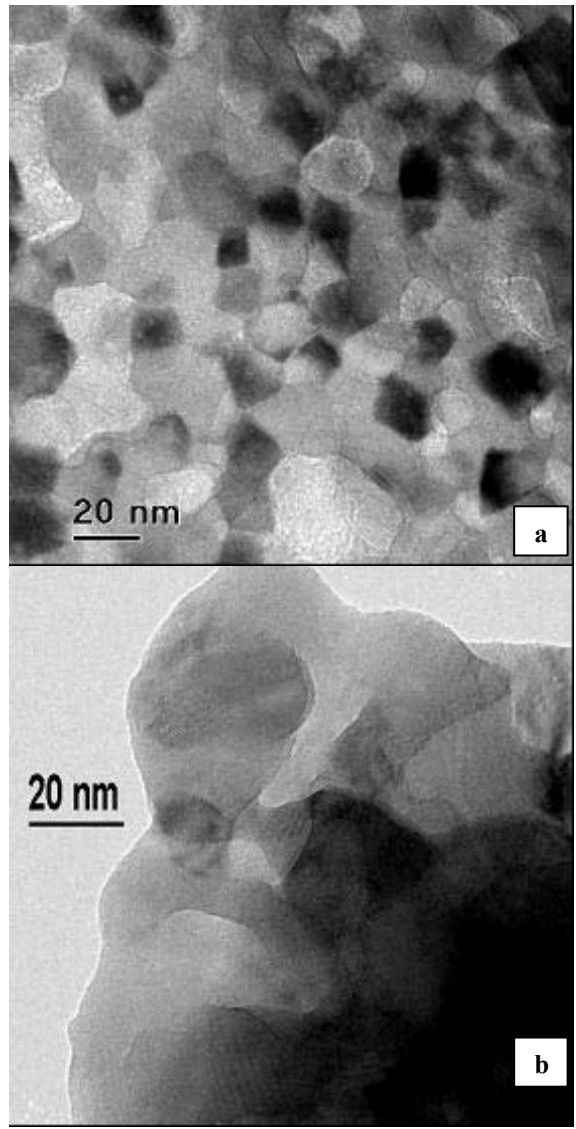

Figure 3

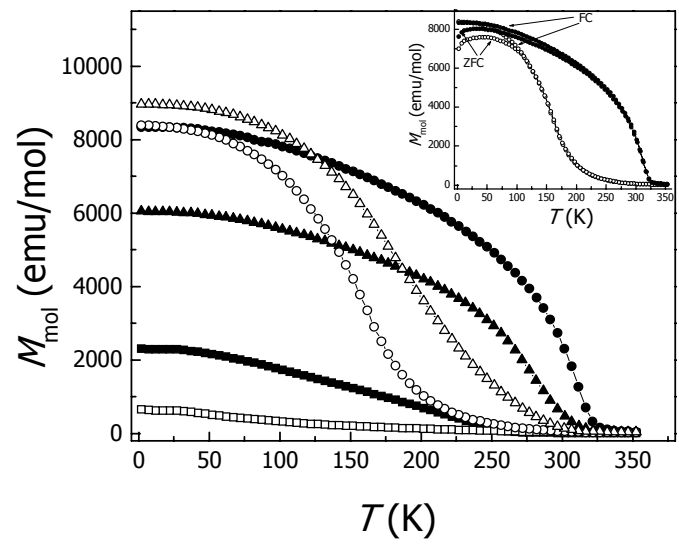

Figure 4

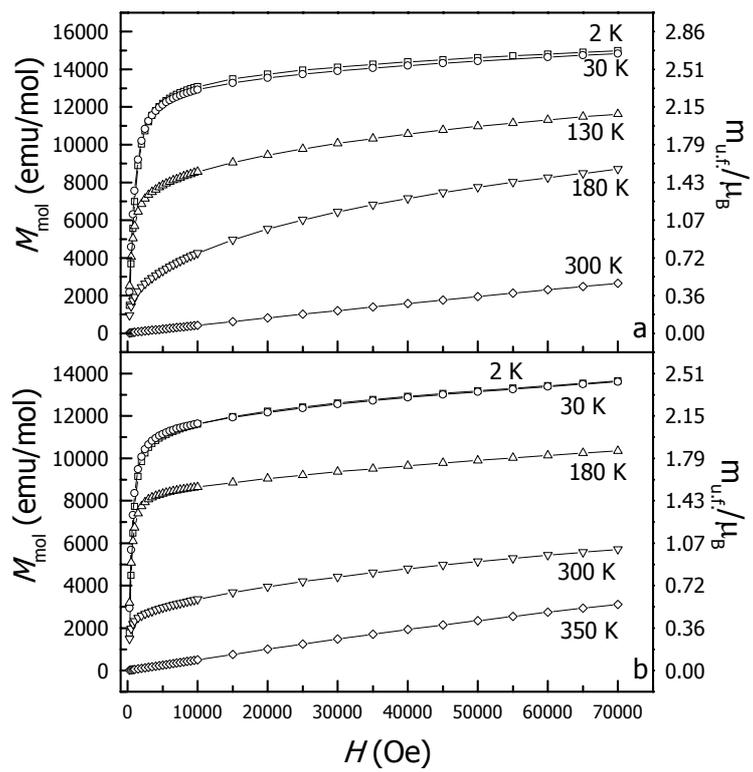

Figure 5

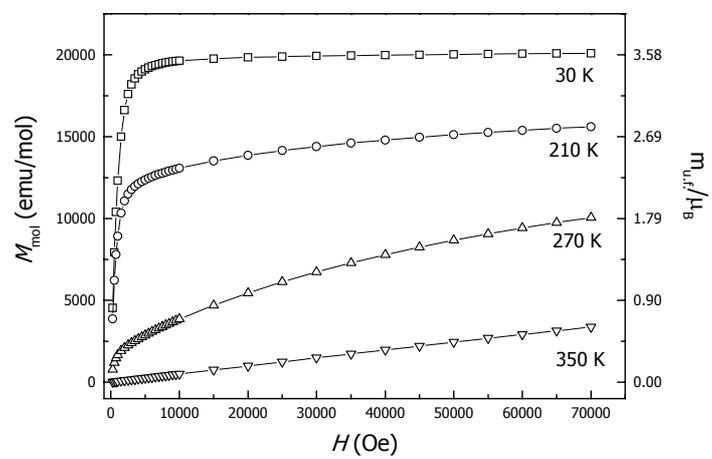

Figure 6

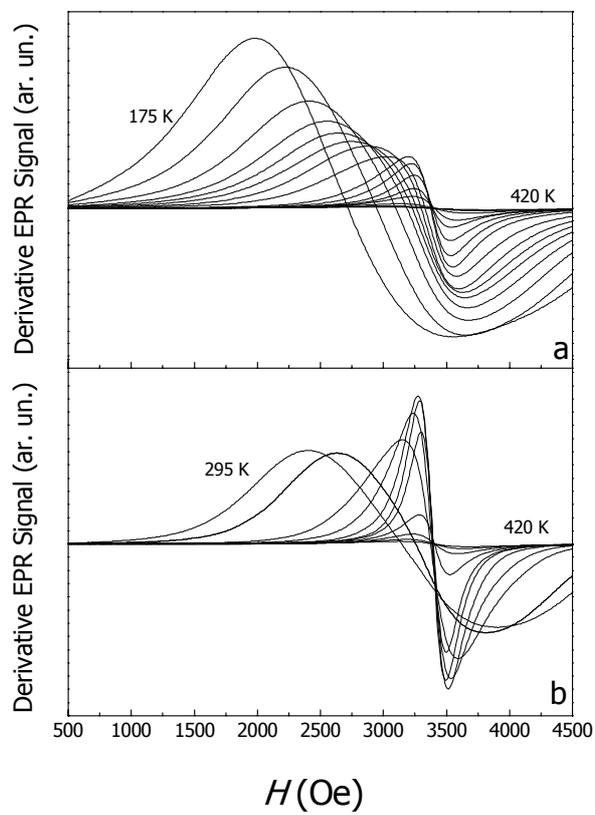

Figure 7

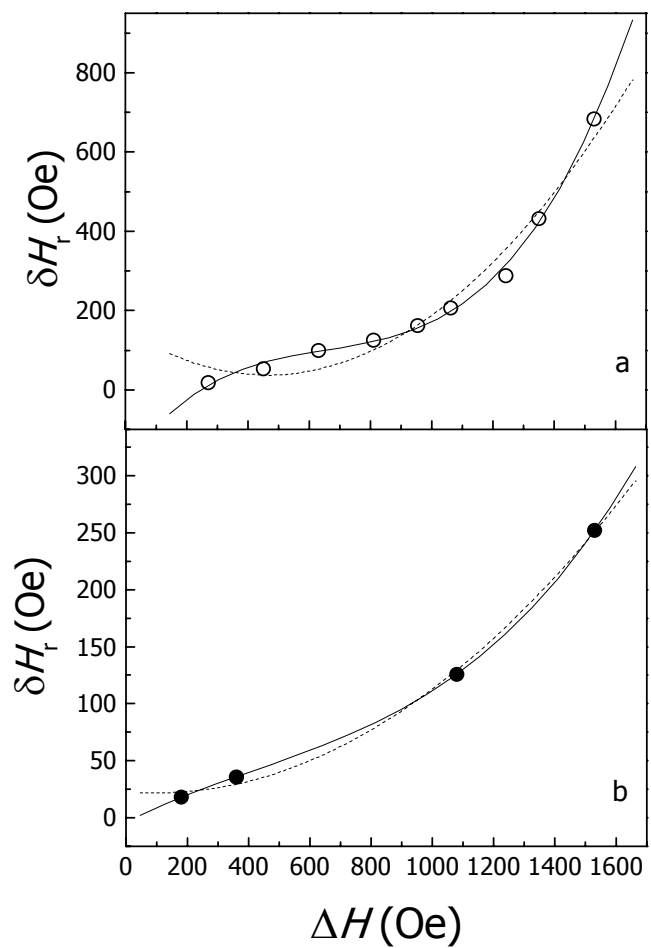

Figure 8

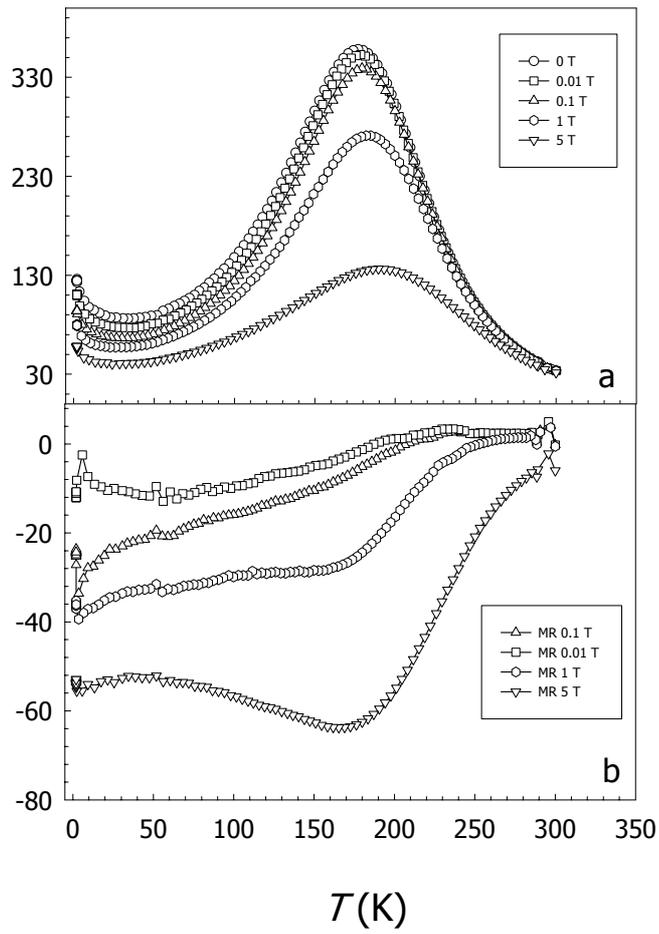

Figure 9

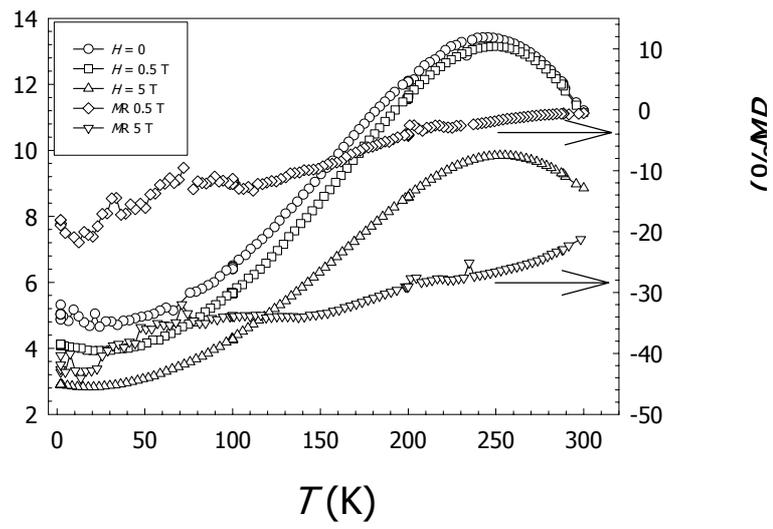

Figure 10